\documentclass{iopart}%[12pt]
\usepackage[dvips]{graphicx}% Include figure files
\begin{document}

\title[Surface-atom force out
of thermal equilibrium and its effect on ultra-cold atoms]{Surface-atom force out
of thermal equilibrium and its effect on ultra-cold atoms}

\author{Mauro Antezza}

\address{Dipartimento di Fisica, Universit\`a di Trento
and CNR-INFM R\&D Center on Bose-Einstein Condensation,
Via Sommarive 14, I-38050 Povo, Trento, Italy}

\ead{antezza@science.unitn.it}

\begin{abstract}
The surface-atom Casimir-Polder-Lifshitz force out of thermal equilibrium is investigated in the framework of  macroscopic electrodynamics. Particular attention is devoted to its large distance limit that shows a new, stronger behaviour with respect to the equilibrium case.
The frequency shift produced by the surface-atom force  on the the center-of-mass oscillations of a harmonically trapped Bose-Einstein condensate and on the Bloch oscillations of an ultra-cold fermionic gas in an optical lattice are discussed for configurations out of thermal equilibrium.
\end{abstract}
%Uncomment for PACS numbers title message
% Keywords required only for MST, PB, PMB, PM, JOA, JOB? 
%\vspace{2pc}
%\noindent{\it Keywords}: Article preparation, IOP journals
% Uncomment for Submitted to journal title message
%\submitto{\JPA}
% Comment out if separate title page not required
%\maketitle
\section{Introduction}
The electromagnetic force felt by a neutral atom near the surface of a substrate has been object of an intense investigation since the pioneering works by Casimir and Polder \cite{CP} and Lifshitz, Dzyaloshinskii and Pitaevskii \cite{Lifshitz_56,LP}. In addition to the fundamental character of the force, these studies \cite{Babb} are presently motivated by the possibility of technological applications \cite{capasso}, by searching stronger constrains on hypothetical non Newtonian forces \cite{Decca} as well as its role in biological systems \cite{Israel}.

New perspectives to study such a force are opened by the recent development in storing and manipulating ultra-cold atoms. Indeed experimental and theoretical research has been recently focused on the forces acting on ultra-cold atomic gases due to the presence of a nearby surface. They include atomic beams \cite{Beams}, Bose-Einstein condensates \cite{articolo1,eric05,bec} and degenerate Fermi gases \cite{Iacopo}.

The surface-atom force at thermal equilibrium $F^{{\rm eq}}(T,z)$ can be in general separated in two parts
\begin{equation}
F^{{\rm eq}}(T,z)=F_{0}(z)+F_{{\rm th}}^{{\rm eq}}(T,z).
\label{Feq}
\end{equation}
The first one, $F_{0}(z)$, is related to zero-point fluctuations ($T=0$) of the
electromagnetic field. At short distances $z$ this force behaves like $1/z^{4}$ and is the analog of the van der
Waals-London inter-atomic force. At larger distances the inclusion of
relativistic retardation effects gives rise to the Casimir-Polder  asymptotic behaviour \cite{CP,LP}
\begin{equation}
F_{0}(z)_{z\to \infty} =
-\frac{3}{2}\frac{\hbar c \alpha_0}{\pi z^5}
\frac{\varepsilon_0-1}{\varepsilon_0+1}\phi(\varepsilon_0),
\label{Feqvacasymp}
\end{equation}
where $\alpha_0$ and $\varepsilon_0$ are the static polarizability of the atom and the
static dielectric function of the substrate respectively. The function $\phi(\varepsilon_0)\sim1$ is defined, for example, in \cite{articolo1}.
 The second contribution to the force, $F_{{\rm th}}^{{\rm eq}}(T,z)$, is due to the  thermal fluctuations of
the electromagnetic field. This contribution was first considered by
Lifshitz \cite{lifshitzDAN} who applied the theory of electromagnetic fluctuations developed by Rytov \cite{Rytov}. At large distances the thermal contribution approaches the so-called Lifshitz law
\begin{equation}
F_{{\rm th}}^{{\rm eq}}(T,z)_{z\to \infty}  =
-\frac{3}{4}\frac{k_{\rm B}T \alpha_0}{z^4}\frac{\varepsilon_0-1}{\varepsilon_0+1}.
\label{neweqasymp}
\end{equation}
Such asymptotic behaviour is reached at distances larger
than the thermal wavelength $\lambda_{\rm T}=\hbar
c/k_{B}T$, corresponding to $\sim 7.6\; \; \mu m$ at room temperature. Thus it is the leading contribution to the total force. 

The Lifshitz force was originally evaluated  at thermodynamic
equilibrium. A non-trivial issue is the study of such a force out of thermal
equilibrium, characterizing configurations
where the temperature of the
substrate $T_{\rm S}$ and  environment $T_{\rm E}$, do not coincide. For instance in typical 
experiments with ultra-cold atomic gases the environment temperature is determined by the chamber containing
the substrate and the trapped atoms.
 
In this paper we describe the surface-atom force out of thermal equilibrium and how to recover its  asymptotic behaviour at large distances. We assume that the radiation surrounding the atom is not able to populate its internal  excited states which are assumed to be at energies
$\hbar \omega_{at}$ much higher than the  thermal energy:
\begin{equation}
k_{\rm B}T_{\rm S}, \; k_{\rm B} T_{\rm E} \ll \hbar \omega_{at}\,.
\label{condition}
\end{equation}
This condition is very well satisfied at ordinary temperatures (for example the first optical resonance of Rb atoms corresponds to $1.8 \; 10^4$K). 
In the last part of the paper we analyze the effects of such a force on cold atoms, and in particular on the center-of-mass motion of a trapped Bose-Einstein condensate and on the Bloch oscillations of ultra-cold fermionic atoms in an optical lattice.  
%%%%%%%%%%%%%%%%%%%%%%%%%%%%%%%%%%%%%%%%%%%%%%%%%%%%%%%%%%%%%%%%%%%%%%%%%%%%%%%%%
\section{\label{sec:SIPE} Green-function formalism}
%%%%%%%%%%%%%%%%%%%%%%%%%%%%%%%%%%%%%%%%%%%%%%%%%%%%%%%%%%%%%%%%%%%%%%%%%%%%%%%%%
In the calculation of the surface-atom force the main ingredient is clearly the electromagnetic field and its sources. The latter, in our approach, are treated as point-like oscillating dipoles. Furthermore it is useful to write the fields using the Green's function formalism, the Green's function being  the solution of the wave equation for a point-like source. Once this solution is known, the solution due to a general source can be obtained by the principle of linear superposition. The dyadic Green function ${\bf\overline{ G}}$ describing the electromagnetic field in surface optics (for isotropic, linear and non-magnetic media) is the solution of the equation
\begin{equation}
\nabla\wedge\nabla\wedge{\bf\overline{ G}}[\omega;{\bf r},{\bf r}']-k^2\varepsilon(\omega;{\bf r}){\bf \overline{G}}[\omega;{\bf r},{\bf r}']=4\pi k^2{\bf \overline{I}}\delta({\bf r}-{\bf r}'),
\label{elweqcfe}
\end{equation}
with the boundary conditions imposed by the geometry of the problem. In previous equation $k=\omega/c$ is the vacuum wavenumber, $\varepsilon(\omega;{\bf r})$ is the dielectric function and  ${\bf \overline{I}}$ is the identity dyad. Equation (\ref{elweqcfe}) comes from the usual wave equation for the Fourier transformed electric field
\begin{equation}
\nabla\wedge\nabla\wedge{\bf E}[\omega;{\bf r}]-k^2\varepsilon(\omega;{\bf r})\;{\bf E}[\omega;{\bf r}]=4\pi k^2{\bf P}[\omega;{\bf r}],
\label{elweq}
\end{equation}
obtained from the macroscopic Maxwell equations
in which the sources are described by the effective electric polarization field ${\bf P}[\omega;{\bf r}]$ related to the electric current by ${\bf J}[\omega;{\bf r}]=-i\omega{\bf P}[\omega;{\bf r}]$.
 The convolution of the solution obtained from Eq.(\ref{elweqcfe}) and the effective electric polarization gives the electric field at the observation point ${\bf r}$ 
\begin{equation}
{\bf E}\left[\omega;{\bf r}\right]=\int {\bf \overline{G}}\left[\omega;{\bf r},{\bf r}'\right]\;\cdot {\bf P}\left[\omega;{\bf r}'\right]\;\rmd{\bf r}'.
\label{lafh}
\end{equation}
%
%where the integration is performed over the volume containing the sources at ${\bf r}'$.
%%%%%%%%%%%%%%%%%%%%%%%%%%%%%%%%%%%%%%%%%%%%%%%%%%%%%%%%%%%%%%%%%%%%%%%%%%%%%%%%%%%%%%%
\section{Surface-atom force }
%%%%%%%%%%%%%%%%%%%%%%%%%%%%%%%%%%%%%%%%%%%%%%%%%%%%%%%%%%%%%%%%%%%%%%%%%%%%%%%%%%%%%%%
Let us consider the atom described by its complex dielectric polarizability function $\alpha(\omega)=\alpha'(\omega)+i\alpha''(\omega)$ in a vacuum half space $V_1$ and placed at a distance $z$ from the surface of the dielectric half space $V_2$. Let us choose  an orthogonal coordinate system with the $xy$ plane coincident with the interface and the $z$ axis such that the dielectric occupies the region with $z<0$ and the vacuum  the region with $z>0$.
The force  acting on a neutral atom without a permanent electric dipole moment is \cite{Henkel1}
\begin{eqnarray}
\fl{\bf F}({\bf r})=\left\langle d_i^{{\rm tot}}(t)\nabla' E_i^{\rm tot}({\bf r}',t)\Big{|}_{{\bf r}}\right\rangle\approx
\left\langle d_i^{{\rm ind}}(t)\nabla' E_i^{{\rm fl}}({\bf r}',t)\Big{|}_{{\bf r}}\right\rangle+\left\langle d_i^{{\rm fl}}(t)\nabla' E_i^{{\rm ind}}({\bf r}',t)\Big{|}_{{\bf r}}\right\rangle,
\label{forcepert}
\end{eqnarray}
where $d_i$'s are the atomic electric dipole components, we have used the Einstein's summation convention for repeated indices and $\nabla'\equiv\nabla_{{\bf r}'}$. In Eq.(\ref{forcepert}) the average is done with respect to the state of the atom and of the field and the lowest order in perturbation theory has been considered. The first term describes the (spontaneous and thermal) field fluctuations correlated with the induced dipole, and the second term involves (spontaneous and thermal) dipole fluctuations correlated to the field they induce.
The induced electric dipole for the atom at the position ${\bf r}$ is 
\begin{equation}
{\bf d}^{{\rm ind}}[\omega]=\alpha(\omega)\;{\bf E}^{{\rm tot}}[\omega;{\bf r}]\approx\alpha(\omega)\;{\bf E}^{{\rm fl}}[\omega;{\bf r}],
\label{diprelmks}
\end{equation}
where ${\bf E}^{{\rm fl}}[\omega;{\bf r}]$ is the fluctuating field,  and now $\alpha(\omega)$ is the atomic polarizability of the atom in an unbounded space. By modeling the atom as a point-like source dipole ${\bf d}(t)= {\bf d}[\omega] \rme^{-i\omega t}$ at ${\bf r}$, the corresponding polarization in the frequency domain is ${\bf P}[\omega,{\bf r}'']={\bf d}[\omega]\;\delta({\bf r}''-{\bf r})$, and the electric field at the position ${\bf r}'$ is
\begin{eqnarray}
{\bf E}^{{\rm ind}}[\omega;{\bf r}']={\bf \overline{G}}[\omega;{\bf r}',{\bf r}]\cdot{\bf d}^{{\rm tot}}[\omega]\approx{\bf \overline{G}}[\omega;{\bf r}',{\bf r}]\cdot{\bf d}^{{\rm fl}}[\omega].
\label{kkso}
\end{eqnarray}
%
%where ${\bf \overline{G}}[\omega;{\bf r}',{\bf r}]$ is the dyadic Green function, solution of the wave equation Eq.(\ref{elweqcfe}).
Using Eq. (\ref{diprelmks}) and (\ref{kkso}), the fluctuating dipole and field contributions to the surface-atom force (\ref{forcepert}) read
\begin{eqnarray}
\fl\left\langle d_i^{{\rm ind}}(t)\nabla' E_j^{{\rm fl}}({\bf r}',t)\Big{|}_{{\bf r}}\right\rangle=\int\int\frac{\rmd\omega}{2\pi}\frac{\rmd\omega'}{2\pi}\;\rme^{-i(\omega-\omega')t}
\alpha(\omega)\;\nabla'\left\langle E_i^{{\rm fl}}[\omega;{\bf r}] E_j^{{\rm fl}\dag}[\omega';{\bf r'}]\right\rangle\Big{|}_{{\bf r}},
\label{ffffhhhwiaa}
\end{eqnarray}
\begin{eqnarray}
\fl\left\langle d_i^{{\rm fl}}(t)\nabla' E_j^{{\rm ind}}({\bf r}',t)\Big{|}_{{\bf r}}\right\rangle=\int\int\frac{\rmd\omega}{2\pi}\frac{\rmd\omega'}{2\pi}\;\rme^{-i(\omega-\omega')t}
\nabla' G_{jk}^*[\omega;{\bf r}',{\bf r}]\Big{|}_{{\bf r}}\;\left\langle d_i^{{\rm fl}}[\omega] d_k^{{\rm fl}\dag}[\omega']\right\rangle,
\label{ffffhhhwiazza}
\end{eqnarray}
where the integrations are over the whole real frequency axis.
%%%%%%%%%%%%%%%%%%%%%%%%%%%%%%%%%%%%%%%%%%%%%%%%%%%%%%%%%%%%%%%%%%%%%%%%%%%%%%%%%%%%%%%%%%%%%%%
\section{Surface-atom force at thermal equilibrium}
%%%%%%%%%%%%%%%%%%%%%%%%%%%%%%%%%%%%%%%%%%%%%%%%%%%%%%%%%%%%%%%%%%%%%%%%%%%%%%%%%%%%%%%%%%%%%%%
At thermal equilibrium, in order to calculate the average values in (\ref{ffffhhhwiaa}) and (\ref{ffffhhhwiazza}), it is possible to use the fluctuation dissipation theorem \cite{Rytov,LLP}. One finds for the fluctuating  dipoles 
\begin{equation}
\left\langle d_i^{{\rm fl}}[\omega]\;d_j^{{\rm fl}\dag}[\omega']\right\rangle=
\frac{4\pi\hbar\;\delta(\omega-\omega')\;\delta_{ij}}{1-\rme^{-\hbar\omega/k_{\rm B}T}}\;\alpha''(\omega),
\label{FDT11111}
\end{equation}
and for the fluctuating fields  
\begin{equation}
\left\langle E_i^{{\rm fl}}[\omega;{\bf r}]\;E_j^{{\rm fl}\dag}[\omega';{\bf r}'] \right\rangle=
\frac{4\pi\hbar\;\delta(\omega-\omega')}{1-\rme^{-\hbar\omega/k_{\rm B}T}}\;{\rm Im}\;G_{ij}[\omega;{\bf r},{\bf r}']
\label{FDT22222}.
\end{equation}
After substituting the previous equalities
%equations (\ref{FDT11111}) and (\ref{FDT22222}) respectively 
into Eq.(\ref{ffffhhhwiaa}) and (\ref{ffffhhhwiazza}) and using the reciprocity theorem $G_{ij}[\omega;{\bf r},{\bf r}']=G_{ji}[\omega;{\bf r}',{\bf r}]$, the surface-atom  force at thermal equilibrium becomes 
\begin{equation}
 \fl F^{{\rm eq}}(T,z)=
\frac{\hbar}{\pi}\int_0^{\infty}\rmd\omega\;\coth\left(\frac{\hbar\omega}{2k_{\rm B}T}\right)
{\rm Im}\;\left[\alpha(\omega)\;\partial_{z}\;G_{ii}[\omega;{\bf r},{\bf r}']\Big{|}_{{\bf r}}\right].
\label{T=0mss}
\end{equation}
Because of the relation $\coth\left(\hbar\omega/2 k_{\rm B} T\right)=1+2\bar{n}\left(\omega/T\right)$, where $\bar{n}\left(\omega/T\right)=(\rme^{\hbar\omega/k_{\rm B}T}-1)^{-1}$ is the Bose factor, one can separate in Eq.(\ref{T=0mss}) the zero-point fluctuations contribution $F_{0}(z)$ from the thermal contribution $F_{{\rm th}}^{{\rm eq}}(T,z)$. The latter term is the sum of two contributions arising from the two terms of Eq.(\ref{forcepert}). The first one is due to the field fluctuations and it is linear in $\alpha'$. The second one arises from the dipole fluctuations and it is linear in $\alpha''$. As long as the condition (\ref{condition}) is valid, the field fluctuations contribution is the leading term in $F_{{\rm th}}^{{\rm eq}}(T,z)$.
%%%%%%%%%%%%%%%%%%%%%%%%%%%%%%%%%%%%%%%%%%%%%%%%%%%
\section{\label{sec:FFOTE} Surface-atom force out of thermal equilibrium}
%%%%%%%%%%%%%%%%%%%%%%%%%%%%%%%%%%%%%%%%%%%%%%%%%%
A first important investigation of the
surface-atom force out of thermal equilibrium was
carried out by Henkel {\it et al.} \cite{Henkel1}
who calculated  the force generated by a dielectric
substrate at finite temperature by  assuming that
the environment  temperature  is zero.
The principal motivation of that paper was the study
of the force at short distances.

In this section we analyze the general case of an atom placed in vacuum at a distance $z$ from the flat surface of a substrate that we assume to be locally at thermal
equilibrium at a temperature $T_{\rm S}$ which can be equal or different from  the environment 
temperature $T_{\rm E}$, the global system being  in or out
of thermal equilibrium respectively, but in a stationary regime \cite{articolo2,articolo3}. 
In this configuration it is relatively easy to describe the radiation produced by the flat substrate,while it is less trivial to describe the radiation coming from the environment. To face this problem we 
use the Lifshitz trick \cite{Lifshitz_56} for which the vacuum half space is assumed to be a dielectric locally at thermal equilibrium with temperature $T_{\rm E}$, by introducing an infinitesimal imaginary part of its dielectric function. Using the fluctuation dissipation theorem and after integrating over an infinite volume the vacuum half space produces a radiation corresponding to the one that in a real systems is generated by the environment walls at $T_{\rm E}$.

We refer to the substrate as to the half space $2$ occupying the volume $V_2$ with $z<0$, with dielectric function $\varepsilon_2(\omega)=\varepsilon_2'(\omega)+i\varepsilon_2''(\omega)$ and in thermal equilibrium at the  temperature $T_{\rm S}$. The vacuum half space $1$ instead occupies the volume $V_1$ with $z>0$  and is characterized by a  dielectric function $\varepsilon_1(\omega)=\varepsilon_1'(\omega)+i\varepsilon_1''(\omega)$ and a temperature $T_{\rm E}$. Only after calculating the electric fields in this configuration we set  $\varepsilon_1(\omega)=1$.\\
As well as for the thermal equilibrium case, the surface-atom force out of thermal equilibrium can be written as 
\begin{equation}
F^{{\rm neq}}(T_{\rm S},T_{\rm E},z)=F_{0}(z)+F_{{\rm th}}^{{\rm neq}}(T_{\rm S},T_{\rm E},z),
\label{Feqnnn}
\end{equation}
where the thermal contribution $F_{{\rm th}}^{{\rm neq}}(T_{\rm S},T_{\rm E},z)$, provided the condition (\ref{condition}) is satisfied, is dominated by the thermal part of the fluctuating fields correlation (\ref{ffffhhhwiaa}) only, as at thermal equilibrium\footnote{It is worth noticing that since zero-point fluctuations are not affected by condition (\ref{condition}), in the calculation of the zero temperature force $F_0(z)$  both dipole zero-point fluctuations (\ref{FDT11111}) and field zero-point fluctuations (\ref{FDT22222}) are needed.}.\\
The physical origin of the electromagnetic field is \cite{Rytov} the random fluctuating polarization field ${\bf P}[\omega;{\bf r}]$, whose correlations, at thermal equilibrium, are described by the fluctuation dissipation theorem 
\begin{equation}
\left\langle P_k[\omega;{\bf r}]P_l^*[\omega';{\bf r}']\right\rangle=\frac{\delta(\omega-\omega')\delta({\bf r}-{\bf r}')\delta_{kl}\;\hbar\;\varepsilon''(\omega)}{1-\rme^{-\hbar\omega/k_{\rm B}T}}.
\label{1fdttssscc}
\end{equation}
Since the correlations of the source polarization field  are \emph{local}, the fluctuations of the sources at different points add up incoherently. Therefore we can assume that in the whole space the correlations of the sources are given by equation (\ref{1fdttssscc}), 
valid for source dipoles in the the two half-spaces \emph{assumed to be locally} at thermal equilibrium at two different temperatures \cite{Polder}. In order to calculate the field correlation function (\ref{ffffhhhwiaa}) we express the electromagnetic field in terms of its source polarization field via Eq. (\ref{lafh}), and using the Eq. (\ref{1fdttssscc}) we write the thermal part of the surface-atom force out of thermal equilibrium as
\begin{equation}
F_{{\rm th}}^{{\rm neq}}(T_{\rm S},T_{\rm E},z)=F_{{\rm th}}^{{\rm neq}}(T_{\rm S},0,z)+F_{{\rm th}}^{{\rm neq}}(0,T_{\rm E},z),
\label{nonthermalequil}
\end{equation}
where the first thermal contribution
\begin{equation}
\fl F_{{\rm th}}^{{\rm neq}}
(T_{\rm S},0,z)=
\frac{\hbar}{2\pi^2}\int_0^{\infty}\rmd\omega\frac{\varepsilon_2''(\omega){\rm Re}\left[\alpha(\omega)
\int_{V_2} \;G_{ik}[\omega;{\bf r},{\bf r}']\partial_zG_{ik}^{*}[\omega;{\bf r},{\bf r}']\;\rmd^3{\bf r}'\right]}{\rme^{\hbar\omega/k_{\rm B}T_{\rm S}}-1}\label{tpcpf1ss2}
\end{equation}
arises from the sources in the substrate $V_2$, while the second one
\begin{equation}
\fl F_{{\rm th}}^{{\rm neq}}
(0,T_{\rm E},z)=
\frac{\hbar}{2\pi^2}\int_0^{\infty}\rmd\omega\frac{\varepsilon_1''(\omega){\rm Re}\left[\alpha(\omega)
\int_{V_1} \;G_{ik}[\omega;{\bf r},{\bf r}']\partial_zG_{ik}^{*}[\omega;{\bf r},{\bf r}']\;\rmd^3{\bf r}'\right]}{\rme^{\hbar\omega/k_{\rm B}T_{\rm E}}-1}
\label{eqnapejc}
\end{equation}
is produced by the sources in the vacuum half space $V_1$ \footnote{The Green function $G_{ik}$ then reduces respectively to its
transmitted component in Eq.(\ref{tpcpf1ss2}) \cite{henkelbis} and to its incident, reflected and local component in Eq.(\ref{eqnapejc}) \cite{articolo3}. 
%In Eq. (\ref{eqnapejc}) we should set $\varepsilon_1''(\omega)\rightarrow 0$ at the end of the calculations.
}. It is possible to show that
 the sum of Eq.(\ref{tpcpf1ss2}) and (\ref{eqnapejc}), at the same temperature, reproduce the thermal part of the force at thermal equilibrium  \cite{articolo3}. Indeed it is possible to apply to the whole space the Green's functions property (see, for example, \cite{Kazarinov}) 
\begin{equation}
\int_{\Omega}\textrm{d}{\bf r}\;\varepsilon''({\bf r},\omega)\;G_{ik}[\omega;{\bf r}_1,{\bf r}]G_{jk}^*[\omega;{\bf r}_2,{\bf r}]=4\pi\textrm{Im}\;G_{ij}[\omega;{\bf r}_1,{\bf r}_2],
\label{magicths}
\end{equation}
where the integration is on the volume $\Omega$ such that on its surface the Green function is zero. Than we can express the complete surface-atom force out of thermal
 equilibrium in the convenient form
\begin{equation}
 F^{{\rm neq}}(T_{\rm S},T_{\rm E},z)= F^{{\rm eq}}(T_{\rm E},z)+ F_{{\rm th}}^{{\rm neq}}(T_{\rm S},0,z)-
F_{{\rm th}}^{{\rm neq}}(T_{\rm E},0,z),
\label{fullnoneq}
\end{equation}
where the equilibrium force $F^{{\rm eq}}(T,z)$
is given by (\ref{Feq}) and
 $F_{{\rm th}}^{{\rm neq}}(T,0,z)$
is defined by Eq.(\ref{tpcpf1ss2}).
Consistently with the assumption (\ref{condition}),
in deriving the thermal part of Eq.(\ref{fullnoneq}) we ignored
terms proportional to the imaginary part of the atomic polarizability. For the same reason the {\it wind} contribution
in Eq.(\ref{tpcpf1ss2}) and (\ref{eqnapejc}), related to $\alpha''$, can be
ignored and the real part $\alpha'(\omega)$, corresponding to the {\it dispersive} contribution,
can be replaced with its  static ($\omega=0$) value $\alpha_0$. In this non-absorbing condition the force  of Eq.(\ref{nonthermalequil})  can be also written in the form  $F^{{\rm neq}}_{{\rm th}}(T_{\rm S},T_{\rm E},z)=4\pi\alpha_0\partial_z U_{\rm El}(T_{\rm E},T_{\rm S},z)$ where $U_{\rm El}=\left\langle {E\bf }(z,t)^2\right\rangle/8\pi$ is the thermal component of the electric energy density at the atom position.\\
After some  lengthy algebra
%\footnote{The \emph{dispersive} component
%of the force (\ref{tpcpf1ss2})
%can be explicitly worked out by introducing the
%Fourier transform  $g_{ik}[\omega;{\bf K},z_a,z_b]$ of the
%Green function $G_{ik}[\omega;{\bf r}_a,{\bf r}_b]$
%where ${\bf K}$ is the component of the
%electromagnetic wave-vector  parallel to  the interface.
%By explicitly expressing the function $g_{ik}$  in terms of the
%transmitted Fresnel coefficients \cite{Sipe}
%and using the procedure described in \cite{henkelbis}.} 
we find for Eq.(\ref{tpcpf1ss2}) the relevant result
\begin{eqnarray}
\fl F_{{\rm th}}^{{\rm neq}}(T,0,z)  =
-\frac{2\sqrt{2}\hbar\alpha_0}{\pi\;c^{4}}\int_{0}^{\infty }\rmd
\omega \frac{\omega ^{4}}{\rme^{\hbar \omega /k_{\rm B}T}-1
}
\int_{1}^{\infty }\rmd q \;q\; \rme^{-2z\sqrt{q^{2}-1}\omega /c}\;\sqrt{q^{2}-1}\nonumber\\
\times\sqrt{|\varepsilon(\omega)-q^2|+(\varepsilon'(\omega)-q^2)}
\left( \frac{1}{\left|\sqrt{%
\varepsilon(\omega )-q^{2}}+\sqrt{1-q^{2}}\right| ^{2}}+ \right. \nonumber\\
 \left. \frac{\left( 2q^2-1 \right)(q^2+
|\varepsilon(\omega)-q^2|)}{\left|\sqrt{\varepsilon(\omega )-q^{2}}%
+\varepsilon(\omega )\sqrt{1-q^{2}}\right| ^{2}}\right),
\label{explicit}
\end{eqnarray}
where we introduced the dimensionless variable $q=K c/\omega$, with $K$ the modulus of the
electromagnetic wave-vector component parallel to the interface, and $\varepsilon(\omega)\equiv\varepsilon_2(\omega)$.
%===================================================================================================
\begin{figure}[ptb]
\begin{center}
\includegraphics[width=0.50\textwidth]{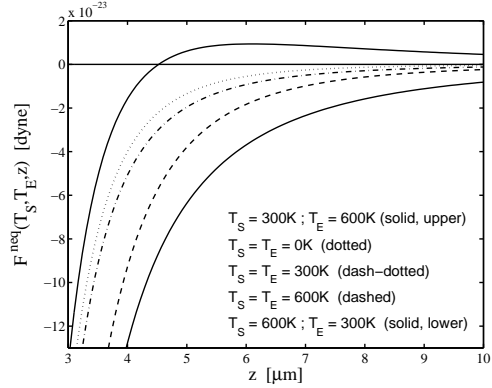}
\caption{\footnotesize Surface-atom force $F^{{\rm neq}}(z)$ 
calculated from Eq.(\ref{fullnoneq}), for different
thermal configurations.}
\label{fig:1}
\end{center}
\end{figure}
%===================================================================================================
In Figure \ref{fig:1}
we show the explicit results for the  full
force obtained from Eq.(\ref{fullnoneq}) as a
function of the distance from the surface for different choices of $T_{\rm S}$ and $T_{\rm E}$.
The calculations have been performed for a sapphire substrate ($\varepsilon_0=9.41$) and for $^{87}$Rb atoms  ($\alpha_0=47.3\;10^{-24}\;$cm$^3$). For $F^{{\rm eq}}(T,z)$ we have used the predictions of \cite{articolo1}. The figure clearly shows that the thermal effects out of equilibrium are sizable (solid lines), thereby providing promising perspectives for future measurements of the surface-atom force at large distances. To increase the attractive nature of the force it is much more convenient to heat the substrate by keeping the environment at room temperature (lower solid line) rather than heating the whole system (dashed line). When $T_{\rm S}<T_{\rm E}$ (upper solid line) the force exhibits a characteristic
change of sign reflecting a repulsive nature at large distances (see also discussion below). At short distances the thermal correction to the force becomes smaller and smaller and is determined by the temperature of the substrate.
The new effects are visible already at distances  $z=4\div 7 \mu m$, where experiments are now becoming available  \cite{eric05}.
%%%%%%%%%%%%%%%%%%%%%%%%%%%%%%%%%%%%%%%%%%%%%%%%%%%%%%%%%%%%%%%%%%%%%%%%%%%%%%%%%%%%%%%%%%%%%
\section{New asymptotic large distance limit}
%%%%%%%%%%%%%%%%%%%%%%%%%%%%%%%%%%%%%%%%%%%%%%%%%%%%%%%%%%%%%%%%%%%%%%%%%%%%%%%%%%%%%%%%%%%%%
In this section we discuss in details the large $z$ behaviour \cite{articolo2} of the out of equilibrium force (\ref{fullnoneq}). After the substitution $q^{2}-1=t^{2}$, Eq.(\ref{explicit}) becomes
\begin{equation}
\fl F_{{\rm th}}^{{\rm neq}}(T,0,z)  =
-\frac{2\sqrt{2}\hbar\alpha_0}{\pi\;c^{4}}\int_{0}^{\infty }\rmd
\omega \frac{\omega ^{4}}{\rme^{\hbar \omega /k_{\rm B}T}-1
}\int_{0}^{\infty }\rmd t\;t^2\;\rme^{-2zt\omega/c}\;f(t,\omega),
\label{oifwc}
\end{equation}
where
\begin{eqnarray}
\fl f(t,\omega) =
\sqrt{|\varepsilon(\omega)-1-t^2|+(\varepsilon'(\omega)-1-t^2)}
\left( \frac{1}{\left|\sqrt{%
\varepsilon(\omega )-1-t^{2}}+it\right| ^{2}}+ \right.\nonumber\\ 
\left. \frac{\left( 2t^2+1 \right)(1+t^2+
|\varepsilon(\omega)-1-t^2|)}{\left|\sqrt{\varepsilon(\omega )-1-t^{2}}%
+i\varepsilon(\omega )t\right| ^{2}}\right).
\label{oifjwc}
\end{eqnarray}
Due to the presence of the exponential $\rme^{-2zt\omega/c}$
 in Eq.(\ref{oifwc}), it is possible to show that only the region $ t\ll 1$ contribute to the large $z$ behaviour  of the force that in such limit 
exhibits the non trivial asymptotic behaviour
\begin{equation}
{F}^{{\rm neq}}_{{\rm th}}(T,0,z)_{z\to \infty}
=-\frac{\sqrt{2}\hbar\alpha_0}{z^{3}2\pi c}%
\int_{0}^{\infty }\rmd\omega \frac{\omega }{\rme^{\hbar \omega /k_{\rm B}T}-1}%
f\left( \omega \right).
\label{LD}
\end{equation}
%
%where we change again variable $u=2zt\omega/c$ and we used that $\int_0^\infty \rmd u\; u^2 \rme^{-u}=2$.
This force exhibits a slower $1/z^3$
decay with respect to  the one holding at thermal
equilibrium where it  decays like $1/z^4$
(see Eq. (\ref{neweqasymp})). In the above equation
we have introduced the low $t$ expansion of Eq.(\ref{oifjwc}) 
\begin{equation}
f\left( \omega \right)=\sqrt{ |\varepsilon(\omega)-1| +
[ \varepsilon^{\prime }(\omega)-1] }\;
\frac{ 2+|\varepsilon(\omega)-1|}{|\varepsilon(\omega)-1|}.
\label{f}
\end{equation}
Result (\ref{LD}) and (\ref{f}) provide the large distance behaviour ($z\rightarrow\infty$) of the force (\ref{explicit}) where the only assumption made was the condition (\ref{condition}). Due to the presence of the Bose factor the force (\ref{LD})  
%and because the small frequency expansion gives a divergent contribution, 
depends on the optical properties of the substrate at frequencies of the order of $\sim k_{\rm B}T/\hbar$.

For temperatures much smaller than $\hbar \omega_c/k_{\rm B}$,   where $\omega_c$ is the lowest
characteristic frequency of the dielectric substrate, only the static value of the dielectric function is relevant and so we can replace $f(t,\omega)$ with its low frequency limit in Eq.(\ref{oifwc}). In this limit $f(t,\omega\rightarrow0)$ is different from zero only for $0<t<\sqrt{\varepsilon_0-1}$, and after the $t\ll1$ expansion Eq.(\ref{oifwc}) becomes
\begin{equation}
\fl {F}^{{\rm neq}}_{{\rm th}}(T,0,z)_{z\to \infty}
=-\frac{\hbar\alpha_0}{z^{3}2\pi c}\frac{\varepsilon_0+1}{\sqrt{\varepsilon_0-1}}
\int_{0}^{\infty }\rmd\omega \frac{\omega }{\rme^{\hbar \omega /k_{\rm B}T}-1}%
\int_0^{2z\sqrt{\varepsilon_0-1}\omega/c} \rmd u\;u^2\rme^{-u},
\label{LDvpk}
\end{equation}
where we performed the change of variable $u=2zt\omega/c$ and replaced
$f(t,\omega\rightarrow0)$ with its $t\ll1$ expansion $\sqrt{2}(\varepsilon_0+1)/\sqrt{\varepsilon_0-1}$.
For 
\begin{equation}
z\gg \frac{\lambda_{\rm T}}{\sqrt{\varepsilon_0-1}},
\label{pfei}
\end{equation} 
where $\lambda_{\rm T}=\hbar\omega/k_{\rm B}T$ is the thermal wavelength, we can extend the upper limit of integration on $u$ to $+\infty$ and so  we obtain that the  force (\ref{fullnoneq}) felt
by the atom approaches the asymptotic behaviour
\begin{equation}
F^{{\rm neq}}(T_{\rm S},T_{\rm E},z)_{z \to \infty} =
-\frac{\pi}{6}\frac{\alpha _{0}k_{\rm B}^2(T_{\rm S}^2-T_{\rm E}^2)}{z^{3}\;
c\hbar}\frac{\varepsilon _{0}+1}{\sqrt{\varepsilon _{0}-1}}.
\label{LevLimit}
\end{equation}
%
%where we used that $\int_0^\infty dx\; x/(\rme^x-1)=\pi^2/6$. 
Result (\ref{LevLimit})
holds  at low temperature  with respect to the first dielectric function resonance ($T\ll\hbar \omega_c/k_{\rm B}$) and at distances satisfying the condition (\ref{pfei}) calculated at the relevant temperatures
 $T_{\rm S}$ and $T_{\rm E}$. Eq.(\ref{LevLimit})
shows that, at large distances, the new force  is
attractive or repulsive depending on whether the
substrate temperature  is higher or smaller than the
environment one. 
 Furthermore, it exhibits a stronger temperature dependence with respect to equilibrium force (\ref{neweqasymp}), contains explicitly the Planck constant and has a $1/z^3$ distance dependence\footnote{Instead of calculating the asymptotic behaviour (\ref{LevLimit}) of the force from the general equation (\ref{explicit}), it is possible to produce a more direct derivation assuming from the very beginning that one can neglect \emph{absorption} and dispersion of the dielectric function of the substrate \cite{LevDirect}.}.
%It is worth to note that the divergence  of the force in the classical limit $\hbar\rightarrow0$ has the same nature if the ultraviolet divergence in the black-body spectrum. 

The new dependence of $F^{{\rm neq}}(T,0,z)$
on temperature and distance can  be physically understood by noticing that the main contribution to the $z$ dependent part of the electric 
energy $U_{\rm El}$ arises from $t\ll1$. Such values of $t$ correspond to the component of the black-body radiation impinging on the surface from the dielectric side in a small interval of angles, of order of $(\lambda_{\rm T}/z)^2$, near the angle of total reflection.  This radiation creates slowly damping evanescent waves in vacuum. As a result $F^{{\rm neq}}(T,0,z)$ turns out to be, in accordance with 
Eq.(\ref{LevLimit}), of order of $-(\alpha_0\lambda_{\rm T}^2/z^3)U_{\rm BB}$, where $U_{\rm BB}\propto T^4$ is the energy density of the black-body radiation.

Equation (\ref{LevLimit}) holds for a dielectric substrate where $\varepsilon_0$ is finite. If we want to find the large distance limit for a metal we should use equation (\ref{LD}). In the limit of small values of $T$
we can use the Drude model. As only frequencies $\omega\sim k_{\rm B}T/\hbar$ contribute, one can substitute in  Eq.(\ref{f})
 $\varepsilon^{\prime \prime }(\omega )
=4\pi \sigma /\omega\gg1$, the real part $\varepsilon^{\prime}(\omega)$
remaining finite as $\omega\to 0$.
 Than one finds $f\left(\omega \right) \rightarrow \sqrt {\varepsilon
^{\prime \prime }(\omega )}=2\sqrt{\pi \sigma/\omega }$, where $\sigma$ is the electric conductivity, so that
 for a Drude metal Eq.(\ref{LevLimit}) is replaced by 
\begin{equation}
F^{{\rm neq}}(T_{\rm S},T_{\rm E},z)_{z\to \infty}=
- \frac{\alpha_0\zeta(3/2)\sqrt{\sigma}\;k_{\rm B}^{3/2}(T_{\rm S}^{3/2}-
T_{\rm E}^{3/2})}{z^{3}\;c\sqrt{2\hbar}},
\label{LevLimMetal}
\end{equation}
where 
%we used that $\int_0^{\infty}\rmd x \;x^{1/2}/(\rme^x-1)=\sqrt{\pi}\zeta(3/2)/2$ with 
$\zeta(3/2)\sim 2.61$ is the Riemann function.  It is easy to show that Eq.(\ref{LevLimMetal}) is valid at the condition
\begin{equation}
z\gg\hbar^{3/2}c \sqrt{4\pi\sigma}/(k_{\rm B}T)^{3/2}.
\label{conlevlim}
\end{equation}
%%%%%%%%%%%%%%%%%%%%%%%%%%%%%%%%%%%%%%%%%%%%%%%%%%%%%%%%%%%%%%%%%%%%%%%%%%%%%%%%%%%%%%%%%%%%%
\section{Effects of the surface-atom force on ultra-cold atoms}
%%%%%%%%%%%%%%%%%%%%%%%%%%%%%%%%%%%%%%%%%%%%%%%%%%%%%%%%%%%%%%%%%%%%%%%%%%%%%%%%%%%%%%%%%%%%%
Ultra-cold gases can provide a useful probe of the surface-atom force. A {\it mechanical} tool sensitive to the gradient of the surface-atom force is in fact the frequency shift of the center-of-mass oscillation of a trapped Bose-Einstein condensate \cite{articolo1,eric05}. On the other hand,  experiments based on Bloch oscillations  are {\it interferometric} tools
sensitive to the force itself \cite{firenze,Iacopo}. Finally one could also think to interference experiments involving 
the macroscopic phase of Bose-Einstein condensates in a double well potential \cite{Ketterle2,Pezze}. The position of the corresponding interference fringes are sensitive to the surface-atom potential.
In the last part of this paper we discuss the first two above mentioned experiments.
%%%%%%%%%%%%%%%%%%%%%%%%%%%%%%%%%%%%%%%%%%%%%%%%%%%%%%%%%%%%%%%%%%%%%%%%%%%%%%%%%%%%%%%%%%%%%
\subsection{Effects on the collective oscillations of a trapped BEC}
%%%%%%%%%%%%%%%%%%%%%%%%%%%%%%%%%%%%%%%%%%%%%%%%%%%%%%%%%%%%%%%%%%%%%%%%%%%%%%%%%%%%%%%%%%%%%
Bose-Einstein condensed gases \cite{rmp} are very dilute, ultra-cold samples characterized by unique properties of coherence and superfluidity. 
%These give rise, among others, to a variety of collective oscillations, i.e. oscillations of the gas density  excited by external perturbations  . 
The study of the collective oscillations \cite{SS} of a Bose-Einstein condensate provides a 
useful probe of the surface-atom potential.  
In fact it is possible to measure with great accuracy the frequency  of the center-of-mass motion $\omega_{\rm CM}$ of a condensate. For a harmonically trapped condensate the frequency $\omega_{\rm CM}$ corresponds to the harmonic trap frequency $\omega_z$, where $z$ is the direction of the oscillations. Thus if a BEC in a harmonic trap is placed at distance $d$ from the surface of a substrate, the surface-atom potential $V_{\rm s-a}(z)$ perturbs the trap potential and produces a shift in $\omega_{\rm CM}$. In the limit of small oscillations (for a complete analysis see \cite{articolo1}), such a shift is
\begin{equation}
\omega_{\rm CM}^2 = \omega_z^2+  \frac{1}{m}\;\int_{-R_z}^{+R_z}\rmd z\;
 n_{0}^{z}(z)\;\partial_z^2V_{\rm s-a}(z+d),
\label{omegaD0}
\end{equation} 
where  $n_0^z(z)$ is $1D$ column density of the gas (density integrated over the directions perpendicular to the direction of oscillation) and $R_z$ is the  Thomas-Fermi radius in the $z$ direction\footnote{For a Bose-Einstein condensate in its ground state  
 the $1D$ column density is easily evaluated in the so called Thomas-Fermi approximation where $n_0^z(z) = 15(1-z^2/R^2_z)^2/16R_z$ \cite{rmp}.}. Therefore  measuring  $\omega_{\rm CM}$ it is possible to extract the surface-atom potential $V_{\rm s-a}(z)$ \cite{articolo1,eric05}.
 
\begin{figure}[ptb]
\begin{center}
\includegraphics[width=0.49\textwidth]{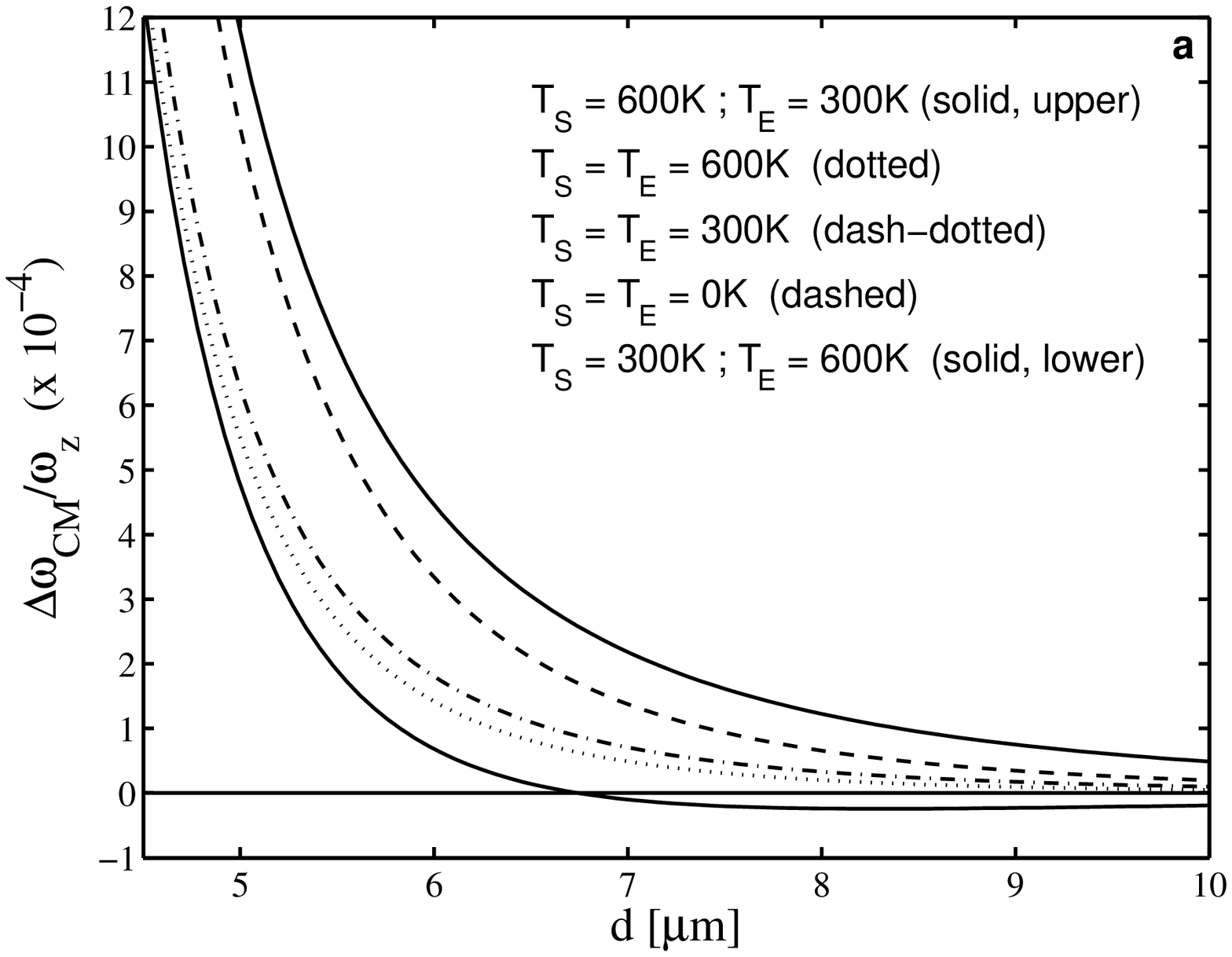}
\includegraphics[width=0.49\textwidth]{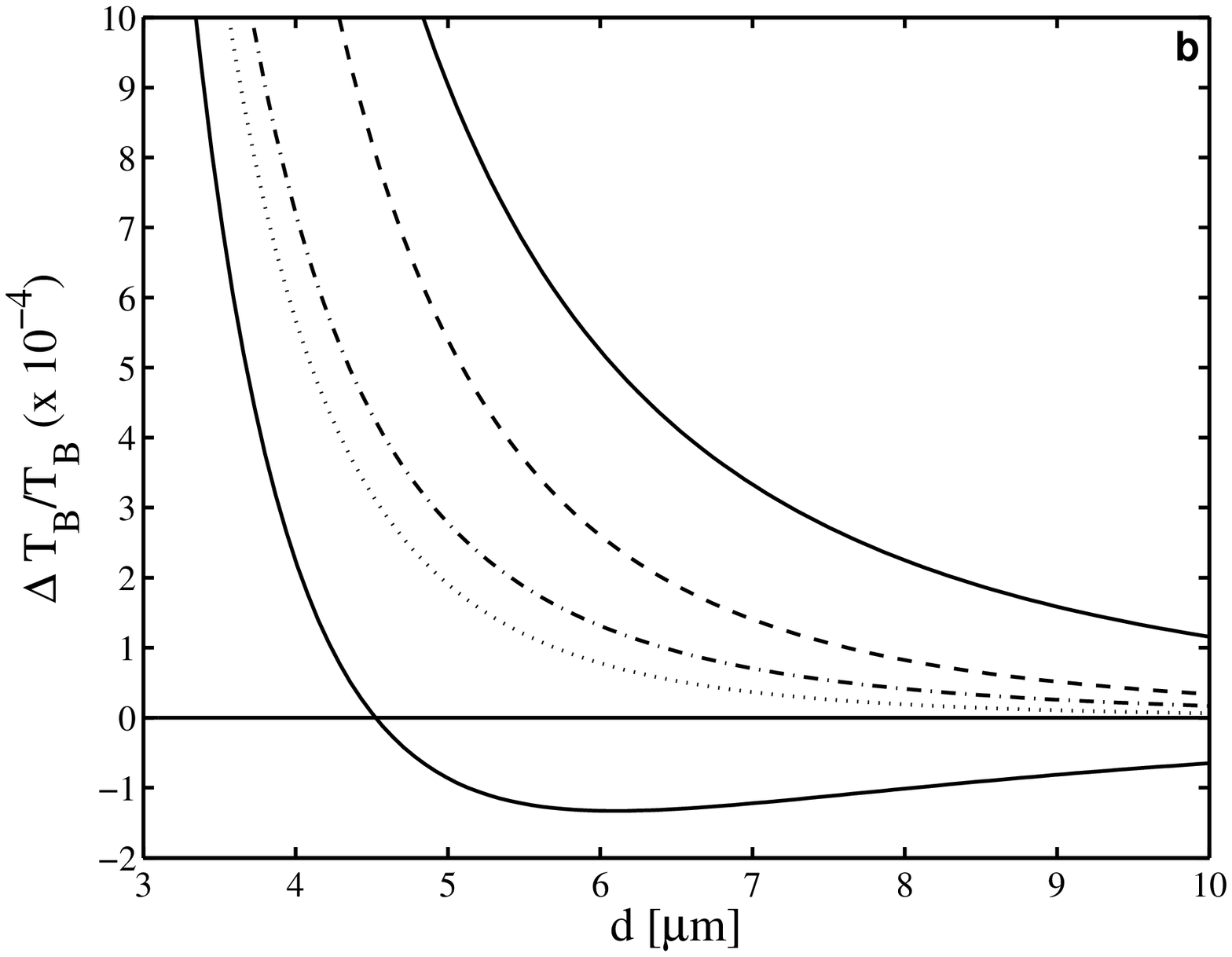}
\caption{\footnotesize ({\bf a}) Relative frequency shift (\ref{omegaD0}) of the center-of-mass oscillation  of a BE condensate ($R_z=2.5\mu m$, $\omega_z/2\pi=220$Hz) and ({\bf b})  relative shift of the Bloch oscillation period of a degenerate Fermi gas (\ref{ojeopi}), out of thermal equilibrium (\ref{Feqnnn}).}
\label{fig:2}
\end{center}
\end{figure}

In fig. \ref{fig:2}({\bf a}) we plotted, as a function of the surface-condensate separation $d$, the relative frequency shift $\Delta\omega_{\rm CM}/\omega_z=(\omega_z-\omega_{\rm CM})/\omega_z$ for the center-of-mass oscillations of a $^{87}$Rb condensate close to a sapphire substrate. In such a calculation we used the surface-atom potential corresponding to the force (\ref{fullnoneq}). 

%===================================================================================================
%===================================================================================================
%%%%%%%%%%%%%%%%%%%%%%%%%%%%%%%%%%%%%%%%%%%%%%%%%%%%%%%%%%%%%%%%%%%%%%%%%%%%%%%%%%%%%%%%%%%%%
\subsection{Effects on Bloch oscillations in Fermi gases}
%%%%%%%%%%%%%%%%%%%%%%%%%%%%%%%%%%%%%%%%%%%%%%%%%%%%%%%%%%%%%%%%%%%%%%%%%%%%%%%%%%%%%%%%%%%%%
When an external force $F_{\rm ext}$ is applied to a particle trapped in a periodical potential, the particle undergoes  oscillations in momentum space (the Bloch oscillations). During this oscillations the particle quasi-momentum $q$ evolves according to $\hbar\dot{q}=F_{\rm ext}$. This is what happens for example in a sample of ultra-cold atoms trapped in a $1D$ optical lattice aligned along the vertical direction.  Bloch oscillations produced by the effect of the gravity force $F_G = mg$  have a period $T_{\rm B}=4\pi\hbar/mg\lambda$ where $\lambda$ is the lattice wave-length and $g$ is gravity acceleration. If now a surface is brought close to the atomic sample the additional surface-atom force $F_{\rm s-a}(z)$ perturbs the gravitational potential and affects the dynamics of the Bloch oscillations (for a complete analysis see \cite{Iacopo,firenze}). In particular it  produces a shift of the period $T_{\rm B}$. In figure \ref{fig:2}({\bf b}) we plotted the relative shift $\Delta T_{\rm B}/T_{\rm B}$ for different thermal configuration as the distance $d$ between the center of a cloud of $^{40}$K fermionic atoms ($\alpha_0=4.3\;10^{-23}$cm$^3$) and the  surface of a sapphire substrate is varied. We used also the approximation of a small cloud of Fermi atoms, for which 
\begin{equation}
\frac{\Delta T_{\rm B}}{T_{\rm B}}=-\frac{F_{\rm s-a}(d)}{mg}.
\label{ojeopi}
\end{equation}
In the range of distances plotted in fig. \ref{fig:2}({\bf b}) this approximation provides results in good agreement with the exact calculation \cite{Iacopo} that takes into account real experimental parameters of the gas.\\
%%%%%%%%%%%%%%%%%%%%%%%%%%%%%%%%%%%%%%%%%%%%%%%%%%%%%%%%%%%%%%%%%%%%%%%%%%%%%%%%%%%%%%%%%%%%%%%%%%
%\section{Conclusions}
%%%%%%%%%%%%%%%%%%%%%%%%%%%%%%%%%%%%%%%%%%%%%%%%%%%%%%%%%%%%%%%%%%%%%%%%%%%%%%%%%%%%%%%%%%%%%%%%%%

%In this paper we investigated the surface-atom force out of thermal equilibrium and its new asymptotic behaviour. The effect of such a force on ultra-cold gases was also described. In particular i
It is worth noticing that both effects of the surface-atom force out of thermal equilibrium described in the last section, and plotted in Figures \ref{fig:2}({\bf a}) and \ref{fig:2}({\bf b}), are in the domain of the present  experimental accuracy \cite{eric05,firenze}.  
%%%%%%%%%%%%%%%%%%%%%%%%%%%%%%%%%%%%%%%%%%%%%%%%%%%%%%%%%%%%%%%%%%%%%%%%%%%%%%%%%%%%%%%%%%%%%%%%%%
\section{Acknowledgment}
%%%%%%%%%%%%%%%%%%%%%%%%%%%%%%%%%%%%%%%%%%%%%%%%%%%%%%%%%%%%%%%%%%%%%%%%%%%%%%%%%%%%%%%%%%%%%%%%%%
New results presented in this paper have been obtained in collaboration with L.P. Pitaevskii and S. Stringari. We are grateful to A. Recati, I. Carusotto, C. Henkel, M.S. Toma$\check{\textrm{s}}$, E. Cornell, J. Obrecht, D.M. Harber, J. McGuirk and S. Reynaud for useful comments. 
%%%%%%%%%%%%%%%%%%%%%%%%%%%%%%%%%%%%%%%%%%%%%%%%%%%%%%%%%%%%%%%%%%%%%%%%%%%%%%%%%%%%%%%%%%%%%%%%%%
\section*{References}
%%%%%%%%%%%%%%%%%%%%%%%%%%%%%%%%%%%%%%%%%%%%%%%%%%%%%%%%%%%%%%%%%%%%%%%%%%%%%%%%%%%%%%%%%%%%%%%%%%


\begin{thebibliography}{10}

\bibitem{CP}  Casimir H B G and Polder D 1948 {\it Phys. Rev.} {\bf 73} 360

\bibitem{Lifshitz_56} Lifshitz E M 1956 {\it Zh. Eksp. Teor. Fiz.} {\bf 29} 94 [1956 {\it Sov. Phys. JETP} {\bf 2} 73]

\bibitem{LP}  Dzyaloshinskii I E, Lifshitz E M and Pitaevskii L P 1961
{\it Advances in Physics} {\bf 38} 165 

\bibitem{Babb} Babb J F {\it et al.} 2004 {\it Phys. Rev. A} {\bf 70} 042901; Caride A O {\it et al.} 2005 {\it Phys. Rev. A} {\bf 71} 042901

\bibitem{capasso}  Chan H B {\it et al.} 2001 {\it Phys. Rev. Lett.} {\bf 87} 211801

\bibitem{Decca} Decca R S {\it et al.} 2005 {\it Annals Phys.} {\bf 318} 37-80 

\bibitem{Israel} Israelachvili J N 1991  {\it Intermolecular and Surface Forces} 2nd Edition (Academic Press, San Diego); Gingell D and Fornes J A 1975 {\it Nature}	{\bf 256} 210


\bibitem{Beams} Sukenik C I {\it et al.} 1993 {\it Phys. Rev. Lett.} {\bf 70} 560; Shimizu F 2001 {\it Phys. Rev. Lett} {\bf 86} 987; H.Oberst {\it et al.} 2005 {\it Phys. Rev. A} {\bf 71} 052901; Druzhinina V and DeKieviet M 2003 {\it Phys. Rev. Lett.} {\bf 91} 193202  

\bibitem{articolo1}  Antezza M, Pitaevskii L P, and Stringari S 2004 {\it Phys. Rev.} A {\bf 70} 053619 

\bibitem{eric05} Harber D M, Obrecht J M, McGuirk J M, and Cornell E A 2005 {\it Phys. Rev.} A {\bf 72} 033610 

\bibitem{bec} Lin Y J {\it et al.} 2004 {\it Phys. Rev.
Lett.} {\bf 92} 050404; Pasquini T A {\it et al.} 2004 {\it  Phys. Rev. Lett.} {\bf 93} 223201

\bibitem{Iacopo}  Carusotto I {\it et al.} 2005 {\it Phys. Rev. Lett.} {\bf 95} 093202

\bibitem{lifshitzDAN}  Lifshitz E M 1955 {\it Doklady Akademii Nauk SSSR} {\bf 100} 879

\bibitem{Rytov} Rytov S M, Kravtsov Y A and Tatarskii  V I  1989 {\it Principles of Statistical Radiophysics, vol.3: Elements of Random Fields}  (Springer, Berlin)

\bibitem{Henkel1}  Henkel C, Joulain K, Mulet J-P, and Greffet J-J 2002 {\it J. Opt. A: Pure Appl. Opt.} {\bf 4} S109 

\bibitem{LLP} Lifshitz  E M and Pitaevskii L P 1991 \emph{Statistical Physics} Part 2 (Pergamon Press, Oxford)

%\bibitem{LLPCM} Landau L D and Lifshitz E M 1960 \emph{Electrodynamics of Continuos Media} (Pergamon Press, Oxford)

\bibitem{articolo2}  Antezza M, Pitaevskii L P, and Stringari S 2005 {\it Phys. Rev. Lett.} {\bf 95} 113202

\bibitem{articolo3}  Antezza M (to be published)

\bibitem{Kazarinov} Henry C H and Kazarinov R F 1996 {\it Rev. Mod. Phys.} {\bf 68} 801

\bibitem{Polder}  Polder D, Van Hove M 1971 {\it Phys. Rev.} B {\bf 4} 3303

%\bibitem{Sipe}  Sipe J E 1987 {\it J. Opt. Soc. Am. B} {\bf 4} 481; Tomas M S 1995 {\it Phys. Rev. A} {\bf 51} 2545 

\bibitem{henkelbis}  Henkel C, Joulain K, Carminati R and Greffet J J 2000 {\it Opt. Commun.} {\bf 186} 57

\bibitem{LevDirect} Pitaevskii L P (submitted to  {\it Jour. of Phys. A}, in this issue).

\bibitem{firenze}  Roati G {\it et al.} 2004 {\it Phys. Rev. Lett.} {\bf 92} 230402 

\bibitem{Ketterle2}  Shin Y {\it et al.} 2004 {\it Phys. Rev. Lett.} {\bf 92} 050405

\bibitem{Pezze} Collins L A {\it et al.} 2005 {\it Phys. Rev. A} {\bf 71} 033628

\bibitem{rmp} Dalfovo F, Giorgini S, Pitaevskii L P, and Stringari S 1999
{\it Rev. Mod. Phys.} {\bf 71} 463

\bibitem{SS}  Stringari S 1996 {\it Phys. Rev. Lett.} {\bf 77} 2360

\end{thebibliography}
\end{document}